\documentclass[12pt]{article}
\catcode`\@=11
\def\marginnote#1{}

\newcount\hour
\newcount\minute
\newtoks\amorpm
\hour=\time\divide\hour by60 \minute=\time{\multiply\hour by60 \global\advance\minute by-\hour}
\edef\standardtime{{\ifnum\hour<12 \global\amorpm={am}%
        \else\global\amorpm={pm}\advance\hour by-12 \fi
        \ifnum\hour=0 \hour=12 \fi
        \number\hour:\ifnum\minute<10 0\fi\number\minute\the\amorpm}}
\edef\militarytime{\number\hour:\ifnum\minute<10 0\fi\number\minute}

\def\draftlabel#1{{\@bsphack\if@filesw {\let\thepage\relax
      \xdef\@gtempa{\write\@auxout{\string
          \newlabel{#1}{{\@currentlabel}{\thepage}}}}}\@gtempa \if@nobreak
    \ifvmode\nobreak\fi\fi\fi\@esphack} \gdef\@eqnlabel{#1}}
    \def\@eqnlabel{}
\def\@vacuum{}
\def\draftmarginnote#1{\marginpar{\raggedright\scriptsize\tt#1}}

\def\draft{
  \oddsidemargin -.5truein
  \def\@oddfoot{\footnotesize \sl preliminary draft \hfil
    \rm\thepage\hfil\sl\today\quad\militarytime}
  \let\@evenfoot\@oddfoot \overfullrule 3pt
    \let\label=\draftlabel
    \let\marginnote=\draftmarginnote
  \def\@eqnnum{(\theequation)\rlap{\kern\marginparsep\tt\@eqnlabel}%
    \global\let\@eqnlabel\@vacuum}

  }

\newtheorem{lemma}{Lemma}

\renewcommand{\theequation}{\thesection.\arabic{equation}}
\newcommand{\newsection}{
\setcounter{equation}{0}
\section}
\def\appendix#1{
\addtocounter{section}{1} \setcounter{equation}{0}
\renewcommand{\thesection}{\Alph{section}}
\section*{Appendix \thesection\protect\indent
#1}
}
\newcommand{\cpict}[3]{
\dimen1=#1\advance\dimen1 by-\hsize\divide\dimen1 by-2 \vtop to #2{ \noindent\hskip\dimen1{\special{em:graph
#3.bmp}} \vfil}\hskip-2cm }

\def\be{\begin{equation}}
\def\ee{\end{equation}}
\def\bea{\begin{eqnarray}}
\def\eea{\end{eqnarray}}

\newcommand{\lp}{\left(}
\newcommand{\rp}{\right)}
\newcommand{\scr}{\scriptstyle}

\newcommand{\om}{\omega}
\newcommand{\al}{\alpha}
\newcommand{\lam}{\lambda}

\newcommand{\dV}{\frac{d}{dV}}
\newcommand{\dmul}{\frac{d\mu_\alpha}{dV}(p)}
\newcommand{\parV}{\frac{\partial}{\partial V}}
\newcommand{\cI}{\oint_{\cal C_{\cal D}}\frac{d\om}{2\pi i}}
\newcommand{\Vp}{V^{\prime}}
\newcommand{\pho}{\phi^{(0)}}

\newcommand{\ty}{{\tilde y}}
\newcommand{\SYM}{\hbox{\,Sym}}
\newcommand{\SYMa}{\hbox{\,Sym}(\not\!\mu_\alpha)}

\begin{document}
\setlength{\unitlength}{1.5mm}

\begin{center}
{\Large Genus one correlation to multi-cut matrix model solutions}\\
\vspace{4mm}
\end{center}
\begin{center}
\vspace{4mm} {L. Chekhov\footnote{E-mail: chekhov@mi.ras.ru.}}\\
\vspace{18pt} {\it Steklov Mathematical Institute,}
\\ {\it  Gubkina 8, 117966, GSP--1, Moscow, Russia}
\end{center}
\vskip 0.9 cm
\begin{abstract}
We calculate genus one corrections to Hermitian one-matrix model solution with
arbitrary number of cuts directly from the loop equation confirming the answer
previously obtained from algebro-geometrical considerations and generalizing it
to the case of arbitrary potentials.
\end{abstract}

\renewcommand{\thefootnote}{\arabic{footnote}}
\setcounter{footnote}{0}

\newsection{Introduction}

Solving matrix models using the loop equation has more than decade long history. An
algorithm for finding genus expansion solutions to the Hermitian one-matrix model
was elaborated (see~\cite{AMB93} and references therein) in the case where the
limiting eigenvalue distribution spans a single interval on the real line in the
large~$N$ limit (in what follows, $N$ is the dimension of the Hermitian matrix).
Such a situation is commonly called the one-cut solution. Results for multi-cut
solutions were few. The problem was that it is not too difficult to find a solution
for the loop equation, i.e., the genus expansion of the so-called loop mean---this
problem was solved by Akemann~\cite{Ak96}. The universal critical behavior of the
corresponding correlators was considered in~\cite{AkAm}.
However, a difficult problem is to
integrate the obtained solution in order to produce higher genus contributions to
the free energy. Akemann has succeeded in finding such a genus-one contribution in
the two-cut case only.

Recently, it was realized that multi-cut solutions to matrix models is the powerful tool
for studying broken vacua structures of ${\cal N}=2$ super Yang--Mills theory of Seiberg
and Witten~\cite{SW}. Simultaneously, a new relation between the superpotentials of
${\cal N}=1$ supersymmetric gauge theories in four dimensions and free energies of matrix
models in the planar limit was proposed \cite{CIV,DV}. The superpotentials in some ${\cal N}=1$
four-dimensional Yang--Mills theories has turned out to be expressed through a single
holomorphic function \cite{CIV}, which was later identified with the planar limit of
the free energy of the multi-support solutions to matrix models~\cite{DV}.

On the other hand, there are strong indications that such an analogy goes beyond the
tree level and just the large-$N$ limit of the matrix model partition function. The
relation between gauge theories and matrix model solutions were checked on the
one-loop level for the solution with two cuts and a cubic matrix model
potential~\cite{KMT}. A comparison in general case must involve the structure of the
general matrix model multi-cut solution in genus one approximation, which has been
missed as yet. Kostov~\cite{Kos} had provided arguments in favor of using the
conformal field theory (CFT) methods for calculating such the corrections in the general
case following the way proposed by Moore~\cite{Moore}; this method
has been improved by Dijkgraaf, Sinkovics, and
Tem\"urhan~\cite{DST} who used algebro-geometrical and CFT arguments for proposing
the formula for genus one correction and comparing it with the gauge theory answer.
Note, however, that the very applicability of Kostov's star operator (dressing)
method needs to be justified by direct calculations using the loop equation alone
and, second, the algebro-geometrical methods need to be improved in a way to
allow calculating contributions from
"non-geometrical" part of the action---from the so-called "moments" of the matrix
model potential (such contributions always arise if the power of the polynomial that
is the derivative of the potential exceeds the number of cuts in play; then, double
points arise and the genus of the arising hyperelliptic surface can be much lesser
than the highest degree of the potential. In this situation, however, the
superpotential satisfying the Whitham hierarchy equations and the WDVV equations is
still attained by the planar (large-$N$) limit of the free energy of the corresponding
matrix model~\cite{ChM},~\cite{ChMMV}.

In this paper, we derive the subleading (genus one) expression for the free energy
of the generalized matrix model multi-cut solution directly from the loop equation,
without referring to geometry or to the CFT reasoning thus supporting the
calculations in~\cite{Kos} and~\cite{DST}. In Secs.~2,~3, we closely follows the
method of~\cite{AMB93},~\cite{Ak96}reviewing the iterative procedure for finding
the loop mean in the genus expansion; the answer for the genus-one
contribution $W_1(p)$ to the loop mean is presented
in Sec.~4. We integrate the obtained answer in Sec.~4
thus obtaining the complete expression for the genus-one part $F_1$ of the free
energy. As a byproduct, we obtain interesting combinatorial identities on the
quantities entering $W_1(p)$.

\newsection{Definitions. Loop equation} \label{basic}

The partition function of the Hermitian one-matrix model is
\be
{\cal Z}[N,\{t_i\}]\equiv e^{N^2F}=e^{\sum_{g=0}^\infty N^{2-2g} F_g[\{t_i\}]}  \equiv
\int d\Phi\, e^{-N\, {\rm tr} V(\Phi)},
\label{Z}
\ee
where the integration is over
Hermitian $N\times N$ matrices $\Phi$. The matrix potential, which we assume to be
polynomial of degree $(m+1)$, is given by the power series
\be
V(x) = \sum_{j=1}^{m+1} \frac{t_j}{j}x^j .
\ee
Deriving the loop equation implies
introducing {\em all} coupling constants $t_j$ into play; a specific potential of
finite order can be inspected by setting the extra couplings to zero in the final
result. The coupling constants $t_j$ are then the sources for polynomial expectation
values with the means defined customarily as
$$
\langle Q(\phi) \rangle  =  \frac{1}{\cal Z} \int d\Phi \, Q(\Phi)\,
                      e^{-N \, {\rm tr} V(\Phi)}.
$$
Introducing the loop insertion operator
\be
\dV(p) \ \equiv \ - \sum_{j=1}^\infty \frac{j}{p^{j+1}} \frac{d}{dt_j} \ ,
\label{dV}
\ee
the one-loop mean $W(p)$ can thus be obtained from the free energy $F$
\be
W(p) \equiv \frac{1}{N} \sum_{k=0}^\infty
             \frac{\langle \mbox{Tr\,}\Phi^k \rangle}{p^{k+1}}
  = \frac{1}{N} \left\langle {\rm tr} \frac{1}{p-\phi} \right\rangle
    = \dV (p) F + \frac{1}{p} . \label{W}
\ee
Analogously, all the multi-loop correlators can be derived by applying $\dV \scr (p)$
to $F$ (or to $W(p)$)
\bea
W(p_1,\ldots,p_n) \ &\equiv& \ N^{n-2} \left\langle
{\rm tr}
       \frac{1}{p_1-\phi}
    \cdots {\rm tr} \frac{1}{p_n-\phi}\right\rangle_{\hbox{conn}}  \nonumber \\
&=&\ \dV(p_n)\dV(p_{n-1})\cdots\dV(p_1) F \ ,
       \  \ n \ge 2 . \label{Wdef}
\eea
Here $\hbox{conn}$ refers to the connected part.
Because the loop correlators and the free energy have the same genus expansion,
\be
W(p_1,\ldots,p_n) = \sum_{g=0}^\infty \frac{1}{N^{2g}} W_g(p_1,\ldots,p_n)
\ , \label{Wg}
\ee
Eq. (\ref{Wdef}) is valid for each genus $g\geq 0$ separately. We assume $W(p)$
to be analytic at infinity with the asymptotic behavior
\be
\lim_{p\to\infty}W(p) \ \sim \ \frac{1}{p} \, , \label{Wass}
\ee
i.e.,
\be
W_g(p)|_{p\to\infty} = \frac{1}{p}\delta_{g,0}+O({1}/{p^2}).
\label{Winf}
\ee

This means that higher $W_g(p)$ must be total derivatives:
\be
\label{total}
W_g(p)=\dV(p)F_g,\quad g\ge 1.
\ee

The loop equation is the {\em exact\/} equation satisfied by the loop mean~\cite{Mak}.
Deriving the loop equation is irrelevant to the
multiple cut structure and relies only on
the invariance of the partition function under a
field redefinition $\Phi \rightarrow \Phi + \epsilon/(p-\Phi)$.
The only effect is that the contour ${\cal C_{\cal D}}$ encircles now a number~$n$ of
disjoint intervals on the real line:
\be
\cI \frac{ V^{\prime}(\om)}{p-\om} W(\om) \ = \
   (W(p))^2 + \frac{1}{N^2}\dV(p)W(p),\ \ p\not\in {\cal D},\label{loop}
\ee
where $V^{\prime}(\om)=\sum_j t_j\om^{j-1}$. We assume that
the density
$\rho_N(\lam)\equiv\frac{1}{N}\langle\sum_{i}^N\delta(\lam-\lam_i)\rangle $
of eigenvalues of the matrices $\Phi$ has a compact support ${\cal D}$
as $N\to\infty$. We let ${\cal D}$ to comprise
$n$ disjoint intervals:
\be
{\cal D} \equiv \bigcup_{i=1}^n [\mu_{2i-1},\mu_{2i}],
\quad \mu_1< \mu_2< \ldots < \mu_{2n}.
\ee
Presenting $W(p)$ in terms of $\rho_N(\lam)$, \ $W(p)=\int d\lam \frac{\rho_N(\lam)}{p-\lam}$,
we find that as $N\to\infty$, $W(p)$ is analytic outside $n$ cuts
on the real axis.

Inserting genus expansion (\ref{Wg}) into loop equation
(\ref{loop}), we obtain
\be
\cI \frac{V^{\prime}(\om)}{p-\om} W_0(\om) \ = \ (W_0(p))^2 \label{plan}
\ee
for genus zero and
\be
(\hat{\cal K}-2W_0(p))W_g(p)  =  \sum_{g\prime=1}^{g-1}W_{g\prime}(p)
          W_{g-g\prime}(p)+\dV(p)W_{g-1}(p), \quad g\ge 1,
          \label{loopg}
\ee
for higher genera, where $\hat{\cal K}$ is a linear integral operator
\be
\hat{\cal K} f(p)  \equiv  \cI \frac{\Vp(\om)}{p-\om}f(\om).
\ee
Given $W_0(p)$, one can then determine $W_g(p)$ for $g\geq 1$ iteratively
genus by genus provided the operator $(\hat{\cal K}-2W_0(p))$
can be inverted uniquely.

We can solve Eq. (\ref{plan}) for the planar solution $W_0(p)$
as follows. Deforming the contour in Eq.~(\ref{plan}) to infinity, we obtain
\be
(W_0(p))^2  = \Vp(p) W_0(p)
 +\oint_{\cal C_{\infty}}\frac{d\om}{2\pi i} \frac{\Vp(\om)}{p-\om}W_0(\om).
\ee
The solution is then formally
\bea
W_0(p) &=& \frac{1}{2}\Vp(p) - \frac{1}{2}\sqrt{(\Vp(p))^2+4P(p)},
\nonumber\\
P(p)   &=& \oint_{\cal C_{\infty}}\frac{d\om}{2\pi i}
      \frac{\Vp(\om)}{p-\om}W_0(\om),
\eea
where the minus sign is chosen in order to fulfill the asymptotic
Eq. (\ref{Winf}) and $P(p)$ is a polynomial of power $m-1$.
If $W_0(p)$ has $n$ cuts in the complex plane, we
propose the ansatz for the square root,
\be
W_0(p) = \frac{1}{2}\lp \Vp(p)-{y}(p)\rp, \label{ansatz}
\ee
where
\be
\label{ty}
y(p)\equiv M(p)\ty(p), \quad \hbox{and} \quad
{\ty}(p)\equiv\sqrt{\prod\nolimits_{i=1}^{2n}(p-\mu_i)}
\ee
(the notation comes from~\cite{ChM}),
and $M(p)$ is assumed to be a polynomial of degree $m-n$, which still has to be
determined. By convention, we set $\ty(p)|_{p\to\infty}\sim p^{n}$, and
$M(p)$ is then
\be
M(p) = \oint_{\cal C_{\infty}} \frac{d\om}{2\pi i}
\frac{\Vp(\om)}{(\om-p)\ty(\om)}.\label{M}
\ee
Inserting this solution in Eq. (\ref{ansatz}) and deforming
the contour back, we obtain the planar one-loop correlator
with an $n$-cut structure,
\be
W_0(p) = \frac{1}{2}\cI \frac{\Vp(\om)}{p-\om}\frac{\ty(p)}{\ty(\om)},\quad p\not\in{\cal D}.
\label{W0}
\ee
Then, the planar eigenvalue density $\rho(\lam)\equiv
\lim_{N\to\infty}\rho_N(\lam)$ has the form
\be
\rho(\lam)=\frac{1}{2\pi i}\lim_{\epsilon\to 0}
         \Big( W_0(\lam-i\epsilon)-W_0(\lam+i\epsilon) \Big)
= \frac{1}{2\pi}\,\hbox{Im\,}y(\lam),\quad \lam \in {\cal D}. \label{rho}
\ee

We still must determine the positions of branching points $\mu_\alpha$, $\alpha=1,\ldots,2n$.
Here, we have two sorts of restrictions. The first set comes from
asymptotic conditions (\ref{Winf}) to be satisfied by ansatz (\ref{W0}), from which
we have
\be
\delta_{k,s}  =  \frac{1}{2}
 \cI \frac{\om^k V^{\prime}(\om)}{\ty(\om)}, \quad k=0,\ldots,n.
 \label{Rand1}
\ee
These conditions provide $n+1$ equations for $2n$ constants
$\mu_\alpha$ (which suffices only if $n=1$).

Another bunch of relations cannot be read from the analytical structure alone.
Introducing the occupation numbers
\be
\label{Si}
S_i=\oint_{A_i}\frac{d\om}{4\pi i}\,M(\om)\ty(\om)\equiv
\oint_{A_i}\frac{d\om}{4\pi i}\,y(\om),
\ee
where $A_i$, $i=1,\dots,n-1$ is the basis of $A$-cycles on the hyperelliptic Riemann surface
$\ty^2=\prod_{\alpha=1}^{2n}(x-\mu_\alpha)$
(we may conveniently choose them to be the first $n-1$ cuts),
we can consider these numbers as {\it independent\/} parameters of the theory in the
Dijkgraaf--Vafa setting, thus imposing the restrictions
\be
\label{dVSi}
\dV(p)S_i=0,\quad i=1,\dots,n-1.
\ee

(Adopting a quite opposite standpoint, which is intrinsic for an original approach
to the matrix models, see~\cite{JU90},~\cite{Ak96}, we must look for the genuine
minimum of the matrix model action with respect to variables $S_i$ as well as with
respect to the times $t_j$. This means that partial derivatives of $F_0$ with
respect to $S_i$ must vanish. One can obtain (see~\cite{JU90},~\cite{DV}), that
these derivatives are differences of chemical potential on disjoint cuts and those
are equal to integrals over {\em dual\/} $B$-cycles on the corresponding
hyperelliptic Riemann surface, that is, this set of conditions reads
\be
\label{Dvdual}
\dV(p)\oint_{B_i}d\om\,y(\om)=0.
\ee
Note that, as we shall see below, from the technical viewpoint, it does not matter
which cycle set, $A$- or $B$-cycles, we take as the basic set. That is, the higher
genus corrections become scheme dependent: choosing
determining conditions (\ref{dVSi}) or (\ref{Dvdual}), we obtain
different expressions for the free energy in the the genus one approximation.
These expressions are however of the same type; the difference will be exactly
in the choice of the corresponding basis cycles.)

\newsection{The iterative procedure}\label{iterate}

\subsection{Moments of the model}

We can determine
higher genus contributions iteratively by inverting
genus expanded loop equation (\ref{loopg}). From $W_g(p)$, all
multi-loop correlators of the same genus can be obtained then simply
by applying the loop insertion operator $\dV \scr (p)$ to it (see Eqs.
(\ref{Wdef}) and (\ref{W})).
As in the one-cut solution \cite{AMB93},
changing variables from coupling constants to moments $M_\alpha^{(k)}$
allows expressing higher genus correlators nonperturbatively
in the coupling constants $t_j$, and these correlators turn out to depend
only on a finite number of the moments
\be
M_\alpha^{(k)}\ \equiv \ \cI \Vp(\om)\phi_\alpha^{(k)}(\om),\quad k=1,2,\dots,
\quad \alpha=1,\ldots,2n,
\label{Momente}
\ee
with
\be
\phi_\alpha^{(k)}(\om) \ \equiv\ \frac{1}{(\om-\mu_\alpha)^k}\,\pho(\om),\qquad
\pho(\om) \equiv \frac{1}{\ty(\om)}.
\ee
Note that expression (\ref{M}) implies
\be
\label{**}
M_\alpha^{(1)}=M(\mu_\alpha).
\ee

\subsection{Determination of the basis}

We now introduce a basis for the operator acting on $W_g(p)$ in Eq. (\ref{loopg}),
\be
(\hat{\cal K}-2W_0(p))\ \chi_\alpha^{(k)}(p) \ \equiv \  \frac{1}{(p-\mu_\alpha)^k},
\quad k=1,2,\dots ,\quad \alpha=1,\ldots,2n\, .\label{Basis}
\ee
Given the r.h.s. of Eq. (\ref{loopg}) to be a fractional rational
function of $p$ having poles at the $\mu_\alpha$ only, $W_g(p)$ must
have the following structure:
\be
W_g(p)  = \sum_{k=1}^{3g-1}\sum_{\alpha=1}^{2n} A_{\alpha,g}^{(k)} \chi_\alpha^{(k)}(p),
\quad g\ge 1.
\label{Wstr}
\ee
Here $A_{\alpha,g}^{(k)}$ are complicated functions of $\mu_\beta$ and the
moments $M_\beta^{(k)}$. As the order of the highest pole in $W_g(p)$
is insensitive to a multi-cut structure, $W_g(p)$ will
depend on at most $2n(3g-1)$ moments, just like the one-cut solution case \cite{AMB93}.

A set of basis functions fulfilling Eq. (\ref{Basis}) is defined by
\be
\tilde{\chi}_\alpha^{(k)}(p) \equiv \frac{1}{M_\alpha^{(1)}}\lp \phi_\alpha^{(k)}(p)-
                   \sum_{r=1}^{k-1}M_\alpha^{(k-r+1)}\tilde{\chi}_\alpha^{(r)}(p) \rp,
                   \quad k=1,2,\dots, \quad \alpha=1,\ldots,2n, \label{chi}
\ee
which can be proved by induction. However, this definition is not unique,
as the kernel of $(\hat{\cal K}-2W_0(p))$ is not empty,
\be
\mbox{Ker}(\hat{\cal K}-2W_0(p)) \ =\ \mbox{Span}\{p^l\pho(p);\
                                 l=0,\ldots, n \} \ .
\ee
This follows from boundary conditions
(\ref{Rand1}). Now, because of the asymptotic behavior
(\ref{Winf}), only terms $\sim {\cal O}(\frac{1}{p^2})$ at large $p$ may be
added to $W_g(p)$ for $g\geq 1$. This requirement reduces the set of
zero modes to be
\be
\mbox{allowed zero modes}: \ \ \{  p^l\pho(p); \ l=0,\ldots,n-2 \} \ .
\label{Nm}
\ee
In the general case, any linear combination of $n-1$
functions can be added to $W_g(p)$. But,
besides its asymptotic behavior, $W_g(p)$ must fulfill Eq. (\ref{total}),
which can be used to fix the basis {\em uniquely}. The $p$-dependence
of $W_g(p)$ must be completely absorbed into derivatives with respect to
$\dV\scr (p)$. This is possible only if the basis functions
$\chi_\alpha^{(k)}(p)$ can be expressed completely in terms of
$\frac{d\mu_\alpha}{dV}\scr (p)$ and $\frac{dM_\alpha^{(k)}}{dV}\scr (p)$ as
functions of $p$. So, $\tilde{\chi}_\alpha^{(k)}(p)$ must be redefined
in Eq. (\ref{chi}) in accordance with this demand, which fixes completely
their zero mode components. The derivatives of branching points
$\mu_\alpha$ and the moments
$M_\alpha^{(k)}$ can be obtained by applying the loop insertion operator
$\dV\scr (p)$ to definitions (\ref{Momente}) and using restrictions (\ref{dVSi})
taking into account that
\be
\dV (p)= \parV (p) + \sum_{\alpha=1}^{2n}\dmul \frac{\partial}{\partial \mu_\alpha}
   + \sum_{\alpha=1}^{2n}\sum_{k=1}^{\infty}\frac{dM_\alpha^{(k)}}{dV}(p)
          \frac{\partial}{\partial M_\alpha^{(k)}},
\label{dVsum}
\ee
Using the identity
\be
\parV (p) \Vp(\om)  = -\frac{1}{(p-\om)^2},
\label{pardV}
\ee
the result for the moments reads
\bea
\frac{dM_\alpha^{(k)}}{dV}(p)&=&
     (k+{1}/{2}) \lp M_\alpha^{(k+1)} \dmul-\phi_\alpha^{(k+1)}(p) \rp\nonumber\\
  &&+\ \frac{1}{2} \sum_{\beta=1 \atop \beta\not= \alpha}^{2n}
      \sum_{l=1}^k \frac{1}{(\mu_\beta-\mu_\alpha)^{k-l+1}} \Big( \phi_\alpha^{(l)}(p)-
          M_\alpha^{(l)} \frac{d\mu_\beta}{dV}(p) \Big)                  \nonumber\\
  &&+\ \frac{1}{2} \sum_{\beta=1 \atop \beta\not= \alpha}^{2n}
        \frac{1}{(\mu_\beta-\mu_\alpha)^k}
           \Big( M_\beta^{(1)}\frac{d\mu_\beta}{dV}(p)-\phi_\beta^{(1)}(p) \Big)
      \ , \nonumber\\
   &&\ \ \alpha=1,\ldots,2n \ , \ k=1,2,\dots .
                                                            \label{dM}
\eea
The quantities $\frac{d\mu_\alpha}{dV}\scr (p)$
are given by the solution of the following set of linear equations,
\bea
0&=& \sum_{\alpha=1}^{2n}\lp \mu_\alpha^k M_\alpha^{(1)}\dmul - p^k\phi_\alpha^{(1)}(p)\rp
      +  2k\ p^{k-1}\pho(p), \ \ k=0,\ldots,n \ , \nonumber\\
0&=& \sum_{\alpha=1}^{2n}\lp M_\alpha^{(1)}\dmul -  \phi_\alpha^{(1)}(p)\rp K_{\alpha,j},
    \quad j=1,\ldots,n-1\ ,  \nonumber\\
&&K_{\alpha,j} \ \equiv \oint_{A_j}d\lam\,\frac{\ty(\lam)}{\lam-\mu_\alpha}\ ,
 \label{dx}
\eea
when applying $\dV\scr (p)$ to (\ref{Rand1}) and using (\ref{dVSi}). While the
first relations can be attained easily, deriving the latter can be performed in
a tricky way first proposed by Akemann. Indeed,
\be
\label{qq}
\dV(p)y(\om)=-\frac12 \sum_{\alpha=1}^{2n}M(\om)\ty(\om)\frac1{\om-\mu_\alpha}
\dmul+\ty(\om)\dV(p)M(\om)
\ee
Calculating the second term, we find from (\ref{M}) that
$$
\dV(p)M(\om)=\oint_{{\cal C}_{\infty,p}}\frac{d\xi}{2\pi i}
\left[\frac{\Vp(\xi)}{(\xi-\om)\ty(\xi)}
\left(\frac12\sum_{\alpha=1}^{2n}\frac1{\xi-\mu_\alpha}\dmul\right)-
\frac{1}{(p-\xi)^2(\xi-\om)\ty(\xi)}\right],
$$
and the first term can be evaluated by the residue at infinity replacing
$\Vp(\xi)$ by $M(\xi)\ty(\xi)$ in this limit, while the second term is given by the
residue at $\xi=p$ (as the integration contour must encircle the infinity and $p$).
The first term then becomes
\be
\label{pp}
\frac12\sum_{\alpha=1}^{2n}\frac{M(\om)-M(\mu_\alpha)}{\om-\mu_\alpha}\,\dmul,
\ee
and it partially cancels the first term in (\ref{qq}), the difference is just the term
proportional to $M(\mu_\alpha)=M_\alpha^{(1)}$, while evaluating the second term
upon substituting it into the integral in (\ref{dVSi}) needs partial integration:
\bea
\frac{\partial}{\partial p}\oint_{A_j}d\om\frac1{p-\om}\frac{\ty(\om)}{\ty(p)}&=&
\frac12\oint_{A_j}d\om\frac1{p-\om}\sum_{\alpha=1}^{2n}\left(\frac{1}{\om-\mu_\alpha}
-\frac{1}{p-\mu_\alpha}\right)\frac{\ty(\om)}{\ty(p)}
\nonumber\\
&=&\frac12\sum_{\alpha=1}^{2n}K_{\alpha,j}\phi_\alpha^{(1)}(p),
\eea
which together with (\ref{qq}) and (\ref{pp}) gives (\ref{dx}).

Linear system of equations (\ref{dx}) implies that the
solution always takes the form
\be
M_\alpha^{(1)}\dmul  =  \phi_\alpha^{(1)}(p) + \sum_{l=0}^{n-2}{\cal L}_{\alpha,l}p^l\pho(p),
\quad \al=1,\ldots,2n.
\label{dxsol}
\ee
The quantities ${\cal L}_{\alpha,l}$,
$\alpha=1,\ldots,2n$, $l=0,\ldots,n-2$, are then determined
by the following set of equations:
\bea
0&=& \sum_{\alpha=1}^{2n}\sum_{l=0}^{n-2}{\cal L}_{\alpha,l}\ \mu_\alpha^k\ p^l  -
     \sum_{\alpha=1}^{2n}\sum_{l=0}^{k-1}\mu_\alpha^{k-l-1}p^l  + 2k\,p^{k-1},
     \quad k=0,\ldots,n, \nonumber\\
0&=& \sum_{\alpha=1}^{2n}\sum_{l=0}^{n-2}{\cal L}_{\alpha,l}\ K_{\alpha,j}p^l,
     \quad j=1,\ldots,n-1.
     \label{alsyst}
\eea
We can redefine basis (\ref{chi}) in terms of total derivatives.
Solving Eq. (\ref{dM}) for $\phi_\alpha^{(k+1)}(p)$ and expressing there
the $p$-dependence through $\dV\scr (p)$-terms plus zero modes, we can
obtain a unique basis inductively from Eq. (\ref{chi})
properly subtracting the corresponding zero modes. We then have
\be
\chi_\alpha^{(k)}(p) \equiv \frac{1}{M_\alpha^{(1)}}\lp
                \phi_\alpha^{(k)}(p)\Big |_{\dV -part}-
         \sum_{r=1}^{k-1}M_\alpha^{(k-r+1)}\chi_\alpha^{(r)}(p) \rp,\quad
         \alpha=1,\ldots,2n,\quad n=1,2,\dots\,.
         \label{chitot}
\ee
Few first basis functions are
\bea
\chi_\alpha^{(1)}(p) &=& \dmul,\quad \alpha=1,\ldots,2n,
\nonumber\\
\chi_\alpha^{(2)}(p) &=& -\frac{2}{3}\dV (p) \log |M_\alpha^{(1)}| -\frac{1}{3}
          \sum_{\beta=1 \atop \beta\neq \alpha}^{2n} \dV(p) \log |\mu_\alpha-\mu_\beta|.
\label{basis2}
\eea

\newsection{Calculations in genus one}

\subsection{Genus one partition function}

Having determined the basis, the loop equation can now be inverted step
by step in genus. For genus $g=1$, Eq. (\ref{loopg}) reads
\be
(\hat{\cal K}-2W_0(p))W_1(p)  =  \dV(p)W_0(p). \label{loop1}
\ee
Given $W_0(p)$ (\ref{W0}) and the loop insertion
operator (\ref{dVsum}), the r.h.s. becomes
\bea
\dV(p)W_0(p)    &=& -\frac{3}{16}\sum_{\alpha=1}^{2n}\frac{1}{(p-\mu_\alpha)^2}
            - \frac{1}{8}\sum_{\alpha,\beta=1 \atop \alpha<\beta}^{2n}
              \frac{1}{(p-\mu_\alpha)(p-\mu_\beta)}
             \nonumber\\
           &&+\ \frac{1}{4}\frac{1}{\pho(p)}\sum_{\alpha=1}^{2n}\frac{1}{p-\mu_\alpha}
               M_\alpha^{(1)} \dmul \nonumber\\
             &=&\ \frac{1}{16}\sum_{\alpha=1}^{2n}\frac{1}{(p-\mu_\alpha)^2}
          - \frac{1}{8}\sum_{\alpha,\beta=1 \atop \alpha<\beta}^{2n}
                 \frac{1}{\mu_\alpha-\mu_\beta}\lp\frac{1}{p-\mu_\alpha}-
                 \frac{1}{p-\mu_\beta}\rp
              \nonumber\\
           &&+ \frac{1}{4}\sum_{\alpha=1}^{2n}\sum_{l=0}^{n-2}
           \frac{{\cal L}_{\alpha,l}\mu_\alpha^l}{p-\mu_\alpha}.
 \label{dW0}
\eea
Here we took into account that regular parts coming from
$\frac{p^l}{p-\mu_\alpha}$, $l=1,\ldots,n-2$, vanish due to Eq. (\ref{alsyst}) in order
for $W_0(p,p)=\dV{\scr (p)}W_0(p)$ to satisfy the correct asymptotic behavior, and
we can just replace $p^l$ by $\mu_\alpha^l$ in numerators of such expressions.
The result for the one-loop correlator of genus one with $n$ cuts can now be
easily obtained using Eq. (\ref{basis2}),
\bea
W_1(p)  &=& \frac{1}{16}\sum_{\alpha=1}^{2n}\chi_\alpha^{(2)}(p)
                  - \frac{1}{8}\sum_{1\leq\alpha<\beta\leq 2n}
           \frac{1}{\mu_\alpha-\mu_\beta}\lp
           \chi_\alpha^{(1)}(p)-\chi_\beta^{(1)}(p)\rp \nonumber\\
            && + \frac{1}{4}\sum_{\alpha=1}^{2n}
                 \sum_{l=0}^{n-2}{\cal L}_{\alpha,l}\mu_\alpha^l\chi_\alpha^{(1)}(p)
                 \nonumber\\
&=& \frac{1}{16}\sum_{\alpha=1}^{2n}\chi_\alpha^{(2)}(p)
                  - \frac{1}{8}\sum_{1\leq\alpha<\beta\leq 2n}
           \frac{1}{\mu_\alpha-\mu_\beta}\lp
           \chi_\alpha^{(1)}(p)-\chi_\beta^{(1)}(p)\rp \nonumber\\
            && + \frac{1}{4}\sum_{\alpha=1}^{2n}
                 \sum_{l=0}^{n-2}{\cal L}_{\alpha,l}\mu_\alpha^l\chi_\alpha^{(1)}(p)
                 \nonumber\\
             &=&\ \frac{1}{16}\sum_{\alpha=1}^{2n}
             \lp-\frac{2}{3}\dV (p) \log |M_\alpha^{(1)}| -\frac{1}{3}
          \sum_{\beta=1 \atop \beta\neq \alpha}^{2n} \dV(p) \log |\mu_\alpha-\mu_\beta|\rp
          \nonumber
          \\
          &&- \frac{1}{8}\sum_{\alpha,\beta=1 \atop \alpha<\beta}^{2n}
                 \frac{1}{\mu_\alpha-\mu_\beta}\lp\dmul-\frac{d\mu_\beta}{dV}(p)\rp
              \nonumber\\
           &&+ \frac{1}{4}\sum_{\alpha=1}^{2n}{\cal L}_{\alpha}(\mu_\alpha)\dmul,
\label{W1}
\eea
where the polynomials ${\cal L}_{\alpha}(p)$ are, by definition,
$\sum_{l=0}^{n-2}{\cal L}_{\alpha,l}p^l$.

We now integrate (\ref{W1}) in order to obtain $F_1$.
While integrating first two terms is easy, the term with zero modes is difficult.
Even before presenting the result, we need more notation. Let us introduce the
quantities
\be
\label{Q}
Q_{j,i}\equiv\oint_{A_j}\frac{\lam^{i-1}}{\ty(\lam)}d\lam,\quad i=1,2,\dots
\ee
(part of them constitute the ${\cal A}$-matrix~\cite{Kos},~\cite{DST}) and the
(polynomial) basis of holomorphic differentials on the hyperelliptic Riemann
surface
\be
\label{hatH}
\widehat H_k(\lam)\equiv \sum_{l=1}^{n-1}\widehat H_{l,k}\lam^{l-1},\quad k=1,\dots,n-1,
\ee
such that
\be
\label{norm}
\oint_{A_j}\frac{\widehat H_k(\lam)}{\ty(\lam)}d\lam=\delta_{k,j},
\ee
whence
\be
\label{inverse}
\sum_{l=1}^{n-1} Q_{j,l}\widehat H_{l,k}=\delta_{j,k}\quad\hbox{for}\quad j,k=1,\dots,n-1.
\ee

The following lemma, to be proven in the next subsection, establishes the relation
between zero modes ${\cal L}_{\alpha,l}$ and $Q_{j,i}$.

\begin{lemma}
For quantities ${\cal L}_{\alpha,l}$ determined by system (\ref{alsyst}) and
the quantities (\ref{Q}) and (\ref{hatH}), we have the relation
\be
\label{main}
\sum_{l=0}^{n-2}{\cal L}_{\alpha,l}\mu_\alpha^l=-\sum_{j=1}^{n-1}
\oint_{A_j}\frac{\widehat H_j(\lam)}{\ty(\lam)(\lam-\mu_\alpha)}d\lam,
\quad \alpha=1,\dots,2n.
\ee
\end{lemma}

It follows from Lemma 1 that the last term in (\ref{W1}) is just
$$
-\frac12\sum_{\alpha=1}^{2n}\frac{\partial}{\partial\mu_\alpha}\lp\log\det_{i,j=1,\dots,n-1}
Q_{j,i}\rp\dmul,
$$
i.e.,
\be
\label{F1}
F_1=-\frac1{24}\log\lp\prod_{\alpha=1}^{2n}M_\alpha^{(1)}\cdot\Delta^{4}\cdot
(\det_{i,j=1,\dots,n-1}Q_{j,i})^{12}\rp,
\ee
where $\Delta=\prod_{1\leq\alpha<\beta\leq 2n}(\mu_\alpha-\mu_\beta)$ is the
Vandermonde determinant. This is our final answer for the genus one partition
function.

\subsection{Proof of the lemma}
Note first that all the quantities $Q_{j,i}$, although being dependent only on
$2n$ moduli $\mu_\al$ are nevertheless algebraically independent for all $j$ and
$i<2n$ (that is, no rational functions of $\mu_\alpha$ and $Q_{j,i}$ with $i<2n$
are identically zero at all $\mu_\alpha$). From the combinatorial point of view,
we can therefore consider all such $Q_{j,i}$ as independent quantities. We also prove
relation (\ref{main}) not for ${\cal L}_{\alpha,l}$ themselves but for special sums
\be
\label{RR}
R_{k,i}\equiv \sum_{\alpha=1}^{2n}\mu_\alpha^k{\cal L}_{\alpha,i}.
\ee
If we calculate all $R_{k,i}$ with $k=0,\dots,2n-1$, we automatically find
all ${\cal L}_{\alpha,i}$ through the inverse Vandermonde transformation.
We can prove (\ref{main}) applying the summation with $\mu_\alpha^k$
to both its parts. Obtaining identities at all $k$ is equivalent to proving the lemma.

For shortening notations, it is convenient to introduce the quantities
\be
\label{T}
T_{l,m}\equiv \sum_{j=1}^{n-1}Q^{-1}_{l,j}Q_{j,m},\quad l=1,\dots, n-1,\quad m=1,2,\dots.
\ee
Obviously, $T_{l,m}=\delta_{l,m}$ for $1\le m\le n-1$.

In the l.h.s. of (\ref{main}), we therefore have
\be
\label{T1}
\sum_{\al=1}^{2n}\mu_\al^k{\cal L}_\al(\mu_\al)=\sum_{i=0}^{n-2}R_{k+i,i},
\ee
while in the r.h.s., we find the expression
\bea
&&-\sum_{\al=1}^{2n}
\sum_{j=1}^{n-1}\oint_{A_j}\frac{\sum_{i=1}^{n-1}Q^{-1}_{i,j}x^{i-1}\mu_\al^k}
{\ty(x)(x-\mu_\al)}dx=
\nonumber\\
&=&
-\sum_{\al=1}^{2n}
\sum_{j=1}^{n-1}\oint_{A_j}\frac{\sum_{i=1}^{n-1}Q^{-1}_{i,j}x^{i-1}(\mu_\al^k-x^k)}
{\ty(x)(x-\mu_\al)}
-\sum_{\al=1}^{2n}
\sum_{j=1}^{n-1}\oint_{A_j}\frac{\sum_{i=1}^{n-1}Q^{-1}_{i,j}x^{i+k-1}}{\ty(x)(x-\mu_\al)}
\nonumber\\
&=&
\sum_{s=0}^{k-1}
\sum_{i=1}^{n-1}T_{i,i+s}\lp\sum_{\al=1}^{2n}\mu_\al^{k-s-1}\rp
-2\sum_{i=1}^{n-1}(i+k-1)T_{i,i+k-1},
\label{T2}
\eea
where we have integrated by part in the second term. Thus, we must prove the coincidence of
(\ref{T1}) and (\ref{T2}) for all $k$.

First, it is straightforward to find from the first relations in (\ref{alsyst}) the
expressions for $R_{k,i}$ with $k=0,\dots,n$:
\be
\label{RR1}
R_{k,i}=\theta(k-1-i)\sum_\alpha\mu_\alpha^{k-i-1}-2k\delta_{k-i-1,0},\quad k=0,1,\dots,n,
\ee
where $\theta(n)=1$ for $n\ge0$ and zero otherwise. The complementary set of relations
follows from the second conditions in (\ref{alsyst}). Let us introduce the standard
symmetric functions
\bea
\SYM_k&\equiv& \sum_{1\leq i_1<i_2<\cdots<i_k\leq 2n}\mu_{i_1}\mu_{i_2}\cdots\mu_{i_k},
\quad k\leq 2n,
\nonumber\\
\SYMa_k&\equiv& \sum_{1\leq i_1<i_2<\cdots<i_k\leq 2n \atop i_s\neq \alpha}
\mu_{i_1}\mu_{i_2}\cdots\mu_{i_k},\quad k\leq 2n-1
\nonumber
\eea
with the obvious recurrence relations
\bea
&&\SYMa_k=\SYM_k-\mu_\alpha\SYMa_{k-1},
\nonumber\\
&&\SYM_0=\SYMa_0=1,\quad \SYM_{-1}=\SYMa_{-1}=0.
\label{rec}
\eea
In particular, we have
\bea
&&\SYMa_k=\SYM_k-\mu_\alpha\SYM_{k-1}+\ldots+(-\mu_\alpha)^{s}\SYM_{k-s}+
(-\mu_\alpha)^{s+1}\SYMa_{k-s-1}
\nonumber\\
&&\quad\hbox{for}\quad s=0,\dots,k,
\label{rec1}
\eea
and
\be
\label{rec2}
\sum_{\al=1}^{2n}\SYMa_s=(2n-s)\SYM_s,\quad 0\leq s\leq 2n.
\ee
We use the transformation
\bea
\frac{\ty(x)}{x-\mu_\alpha}&=&\frac1{\ty(x)}{\prod_{\beta\neq\alpha}^{2n}(x-\mu_\beta)}
=\frac1{\ty(x)}{\sum_{l=0}^{2n-1}x^{2n-l-1}(-1)^l\SYMa_l}
\nonumber\\
&=&\frac1{\ty(x)}{\sum_{l=0}^{2n-1}\sum_{s=0}^lx^{2n-l-1}\mu_\alpha^{l-s}(-1)^s\SYM_s}
\label{sym1}
\eea
to obtain the set of constraints
\be
\label{R2}
\sum_{k=0}^{2n-2}\sum_{s=1}^{2n-k-1}Q_{j,2n-k-s}R_{s,i}\SYM_k(-1)^k=0.
\ee
Using the above formulas and contracting the result with $Q^{-1}_{p,j}$, \ $p=1,\dots,n-1$,
we obtain after some algebra using (\ref{RR1}), (\ref{rec1}), and (\ref{rec2})
\bea
0&=&\sum_{\al=1}^{2n}\mu_\al^{n-i}(-1)^{p+n}\SYMa_{n-p-1}+(p-i-1)(-1)^{p-i-1}\SYM_{2n-1-p-i}
\nonumber\\
&&+\sum_{s=0}^{n-p-1}R_{n+s+1,i}(-1)^{p+n+s+1}\SYM_{n-1-p-s}
\nonumber\\
&&+\sum_{r=n}^{2n-1}T_{p,r}(-1)^{r+i+1}(r-1-i)\SYM_{2n-1-r-i},
\quad p=1,\dots,n-1.
\label{R3}
\eea
We can now find all the remaining quantities $R_{n+1+s,i}$ for $s=0,\dots,n-2$ by
inverting the lower triangular matrix standing by these quantities in the second row
of (\ref{R3}). Namely, we introduce the quantities $q_r$, \ $r=0,1,\dots$, using the
determining relation
\be
\label{R4}
\sum_{r=0}^pq_r\SYM_{p-r}(-1)^{p-r}=\delta_{p,0}\quad\hbox{for}\quad p=0,1,\dots.
\ee
The first few $q_r$ are $q_0=1$, \ $q_1=\SYM_1$, \
$q_2=(\SYM_1)^2-\SYM_2$, etc. Except (\ref{R4}), we need another combinatorial
identity on $q_r$, which follows from (\ref{R4}) and (\ref{rec}). Let us multiply
(\ref{R4}) by $\SYM_1\equiv\sum_{\al=1}^{2n}\mu_\al$:
\bea
\delta_{p,0}&=&\sum_{s=0}^p(s+1)\SYM_{s+1}(-)^sq_{p-s}+\sum_{\al=1}^{2n}\mu_\al^2
\sum_{s=0}^p\SYMa_{s-1}(-)^sq_{p-s}
\nonumber\\
&=&\sum_{s=0}^p(s+1)\SYM_{s+1}(-)^sq_{p-s}-\sum_{\al=1}^{2n}\sum_{q=2}^{p+1}\mu_\al^q
\left(\sum_{s=0}^{p+1-q}\SYM_s(-)^sq_{p+1-q-s}\right)
\nonumber\\
&=&\sum_{s=0}^p(s+1)\SYM_{s+1}(-)^sq_{p-s}-\sum_{\al=1}^{2n}\mu_\al^{p+1},
\nonumber
\eea
that is, we obtain the formula
\be
\label{R5}
\sum_{s=0}^{p+1}s\cdot\SYM_s(-1)^sq_{p+1-s}=-\sum_{\al=1}^{2n}\mu_\al^{p+1},\quad
p=0,1,\dots,
\ee
while this sum vanishes for $p=-1$.

From (\ref{R3}), (\ref{R4}), we obtain for $R_{p,i}$ with $p>n$:
\bea
&&R_{n+1+k,i}=-\sum_{r=0}^{k}q_r\sum_{s=1+i}^{n}T_{2n-s,2n-1-k+r}(2n-s-i-1)
(-1)^{s-i-1}\SYM_{s-i-1}
\nonumber\\
&&\ +\sum_{\al=1}^{2n}\mu_\al^{n+k-i}-\sum_{r=0}^{k}q_r(-1)^{i+k-r+n}\SYM_{n+k-r-i}(n-2-k+r-1),
\nonumber\\
&&\quad\quad k=0,\dots,n-2.
\label{R6}
\eea
Two last terms in (\ref{R6}) pertain to a ``regular'' part of formulas (\ref{T1}) and
(\ref{T2}), i.e., to their parts that do not contain $T_{p,i}$ with $p>n-1$. It is
straightforward to see that these regular parts matches. The only nontrivial thing to do is
to check matching ``nonregular'' parts. To make presentation shorter, we use ellipses to
denote omitted ``regular'' parts. We have for $k=0,\dots, n-2$:
\bea
&&\sum_{i=0}^{n-2}R_{i+k,i}=\sum_{i=0}^{n-k}\dots \ +\sum_{p=0}^{k-3}R_{n+1+p,n+1+p-k}
\nonumber\\
&&\ =(\dots)+\sum_{p=0}^{k-3}\left( -\sum_{r=0}^{p}q_r\,\sum_{s=n+2+p-k}^{n}T_{2n-s,2n-1-p+r}
\right.\times
\nonumber\\
&&\quad\quad\times\biggl.
(2n-s-(n+1+p-k)-1)(-1)^{s-n+p+k}\SYM_{s-n-1-p+k-1}\biggr)
\nonumber\\
&&\ =(\dots)-\sum_{0\le r\le s\le k-2}T_{n-2+k-s,2n-1-r}\lp\sum_{p=r}^s(2k-p-s-4)(-)^{s-p}
\SYM_{s-p}q_{p-r}\rp
\nonumber\\
&&\ =(\dots)-\sum_{0\le r\le s\le k-2}T_{n-2+k-s,2n-1-r}\lp[2k-2s-4]\delta_{r,s}+
\lp-\sum_{\al=1}^{2n}\mu_\al^{s-r}+2n\delta_{r,s}\rp\rp
\nonumber\\
&&\ =(\dots)+\sum_{0\le r\le s\le k-2}T_{n-2+k-s,2n-1-r}\sum_{\al=1}^{2n}\mu_\al^{s-r}
\nonumber\\
&&\quad\quad
-\sum_{0\le r\le k-2}2[n+k-r-2]T_{n-2+k-r,2n-1-r},
\label{R7}
\eea
which exactly gives the ``nonregular'' part in expression (\ref{T2}). The lemma is therefore
proved.

\newsection{Conclusion}
The obtained answer (\ref{F1}) has a nice geometrical interpretation in terms of $G$-functions and
isomonodromic deformations (see~\cite{DST},~\cite{DZ},~\cite{KK}). Worth mentioning is the relation
between this answer and the answer in planar limit, which is a superpotential. These geometrical
relations deserve special studying. We hope to provide also a geometrical explanation for
combinatorial identities of type (\ref{main}).

Worth mentioning is what will be the answer for $F_1$ if we assume the ``old'' matrix model approach
with equated chemical potentials on the intervals of eigenvalue distribution. It is easy to see that
the corresponding answer has form (\ref{F1}) with the only difference that instead of integrating
over $A$-cycles in the matrix $Q_{i,j}$, one should integrate over the $B$-cycles. In the particular case
of two-cut solution, this reproduces the Akemann answer.

Another question pertains to the $S$-dependence of $F_1$. The loop equation by definition may fix only the
part of $F_1$ that depends on the potential; in principle, there can be a part that depends only on the
occupation numbers and which cannot be attained by using the Virasoro invariance of the matrix model integral;
however, this part must be saturated by the planar limit of the matrix model integral. It is nevertheless
interesting and important to calculate $S$-derivatives of the answer $F_1$.

The author thanks B. Dubrovin, T. Grava, D. Korotkin, A. Marshakov, A. Mironov, and A. Zabrodin for
numerous discussions on matrix models and related geometry. The author is grateful to the SISSA for the
hospitality during his visit when this paper was initiated.

The paper was partially financially supported by the Russian Foundation for Basic Research
(Grants Nos. 02--01--00484 and NSh-2052.2003.1) and by the Program Nonlinear Dynamics and Solitons.


\newcommand{\NP}[3]{{\sl Nucl. Phys. }{\bf B#1} (#2) #3}
\newcommand{\PL}[3]{{\sl Phys. Lett. }{\bf B#1} (#2) #3}
\newcommand{\PR}[3]{{\sl Phys. Rev. }{\bf #1} (#2) #3}
\newcommand{\IMP}[3]{{\sl Int. J. Mod. Phys }{\bf #1} (#2) #3}
\newcommand{\MPL}[3]{{\sl Mod. Phys. Lett. }{\bf A#1} (#2) #3}

\end{document}